# Momentum representation for equilibrium reduced density matrices

V. A. Golovko

Moscow State Evening Metallurgical Institute

Lefortovsky Val 26, Moscow 111250, Russia

E-mail: mgvmi-mail@mtu-net.ru

Abstract

The hierarchy of equations for reduced density matrices that describes a thermodynamically equilibrium quantum system obtained earlier by the author is investigated in the momentum representation. In the paper it is shown that the use of the momentum representation opens up new opportunities in studies of macroscopic quantum systems both nonsuperfluid and superfluid. It is found that the distribution over momenta in a quantum fluid is not a Bose or Fermi distribution even in the limit of practically noninteracting particles. The distribution looks like a Maxwellian one although, strictly speaking, it is not Maxwellian. The momentum distribution in a quantum crystal depends upon the interaction potential and the crystalline structure. The momentum distribution in a superfluid contains a delta function. The momentum distribution for the condensate in a superfluid crystal consists of delta peaks that are arranged periodically in momentum space. The periodical structure remains if the condensate crystal is not superfluid.

## 1. Introduction

In Ref. [1], a hierarchy of equations for equilibrium reduced density matrices was obtained, the hierarchy going over, in the classical limit, into the well-known equilibrium Bogolyubov-Born-Green-Kirkwood-Yvon (BBGKY) hierarchy (see also [2] where a systematic exposition of the approach used is presented and various results achieved with its help are discussed). The hierarchy was written down in the coordinate (position) representation. At the same time, some problems of statistical mechanics require momentum space. The aim in the present paper is to investigate the equilibrium reduced density matrices in the momentum representation. It will be shown that this sheds new light on properties of macroscopic quantum systems both nonsuperfluid and superfluid.

## 2. Momentum representation

We consider a system of $N$ particles enclosed in a volume $V$. The particles of mass $m$ interact via a two-body potential $K(|\mathbf{r}_j - \mathbf{r}_k|)$. For the sake of simplicity we shall restrict ourselves to the case of spinless bosons ($^4$He atoms, for example). If need be, the spin can be taken into account analogously with Ref. [3].

Let $\Psi(\mathbf{x}_N, t)$ be the wavefunction of the system. Here and in the following we use the abbreviated notation

$$\mathbf{x}_s = \mathbf{r}_1,\ldots,\mathbf{r}_s;\quad \mathbf{m}_s = \mathbf{p}_1,\ldots,\mathbf{p}_s;\quad d\mathbf{x}_s = d\mathbf{r}_1\cdots d\mathbf{r}_s;\quad d\mathbf{m}_s = d\mathbf{p}_1\cdots d\mathbf{p}_s;\quad s = 1,\ldots, N. \tag{2.1}$$

In the present paper we shall imply pure states alone because the use of mixed states in the problem under study but complicates the notation without giving any advantage [1,2]. For a pure state, the wavefunction depends upon the time $t$ even at thermodynamic equilibrium owing to time-dependent fluctuations. For brevity, we shall omit the argument $t$; all the more so because we shall consider that part in reduced density matrices which does not depend on the time at thermodynamic equilibrium and which describes equilibrium properties [1,2].

The wavefunction in the momentum representation is obtained by Fourier transformation [4]:

$$\widetilde{\Psi}(\mathbf{m}_N) = \frac{1}{(2\pi\hbar)^{3N/2}} \int \Psi(\mathbf{x}_N)\exp\left(-\frac{i}{\hbar}\sum_{k=1}^{N}\mathbf{p}_k\mathbf{r}_k\right)d\mathbf{x}_N, \tag{2.2}$$

$$\Psi(\mathbf{x}_N) = \frac{1}{(2\pi\hbar)^{3N/2}} \int \widetilde{\Psi}(\mathbf{m}_N)\exp\left(\frac{i}{\hbar}\sum_{k=1}^{N}\mathbf{p}_k\mathbf{r}_k\right)d\mathbf{m}_N. \tag{2.3}$$

Herefrom, functions relevant to momentum space are marked off by a tilde.

We define the density matrix $R_N$ and the $s$-particle reduced density matrices in the position representation by [1,2]

$$R_s(\mathbf{x}_s, \mathbf{x}'_s) = \frac{N!}{(N-s)!} \int \Psi(\mathbf{x}_s, \mathbf{r}_{s+1},...,\mathbf{r}_N) \Psi^*(\mathbf{x}'_s, \mathbf{r}_{s+1},...,\mathbf{r}_N) d\mathbf{r}_{s+1} \cdots d\mathbf{r}_N, \; s=1,...,N. \quad (2.4)$$

Analogously to (2.4), we define the reduced density matrices in the momentum representation as

$$\widetilde{R}_s(\mathbf{m}_s, \mathbf{m}'_s) = \frac{N!}{(N-s)!} \int \widetilde{\Psi}(\mathbf{m}_s, \mathbf{p}_{s+1},...,\mathbf{p}_N) \widetilde{\Psi}^*(\mathbf{m}'_s, \mathbf{p}_{s+1},...,\mathbf{p}_N) d\mathbf{p}_{s+1} \cdots d\mathbf{p}_N. \quad (2.5)$$

In order to obtain a relation between $R_s(\mathbf{x}_s, \mathbf{x}'_s)$ and $\widetilde{R}_s(\mathbf{m}_s, \mathbf{m}'_s)$ we substitute (2.3) into (2.4) and utilize the following representation of the Dirac delta function [4]

$$\delta(\mathbf{r}) = \frac{1}{(2\pi\hbar)^3} \int_{(\infty)} e^{\frac{i}{\hbar}\mathbf{p}\mathbf{r}} d\mathbf{p}, \qquad \delta(\mathbf{p}) = \frac{1}{(2\pi\hbar)^3} \int_{(\infty)} e^{\frac{i}{\hbar}\mathbf{p}\mathbf{r}} d\mathbf{r}. \quad (2.6)$$

It will be noted that, if one integrates over the volume $V$ in the last integral, one obtains $\delta(0) = V/(2\pi\hbar)^3$ instead of $\delta(0) = \infty$. We imply, however, that $V \to \infty$. This will permit us to put $\delta^2(\mathbf{p}) = V\delta(\mathbf{p})/(2\pi\hbar)^3$ in some subsequent calculations.

The above substitution of (2.3) into (2.4) yields

$$R_s(\mathbf{x}_s, \mathbf{x}'_s) = \frac{1}{(2\pi\hbar)^{3s}} \int \widetilde{R}_s(\mathbf{m}_s, \mathbf{m}'_s) \exp\left[\frac{i}{\hbar} \sum_{k=1}^{N} (\mathbf{p}_k \mathbf{r}_k - \mathbf{p}'_k \mathbf{r}'_k)\right] d\mathbf{m}_s d\mathbf{m}'_s, \quad (2.7)$$

$$\widetilde{R}_s(\mathbf{m}_s, \mathbf{m}'_s) = \frac{1}{(2\pi\hbar)^{3s}} \int R_s(\mathbf{x}_s, \mathbf{x}'_s) \exp\left[-\frac{i}{\hbar} \sum_{k=1}^{N} (\mathbf{p}_k \mathbf{r}_k - \mathbf{p}'_k \mathbf{r}'_k)\right] d\mathbf{x}_s d\mathbf{x}'_s. \quad (2.8)$$

We introduce also diagonal elements of the density matrices:

$$\rho_s(\mathbf{x}_s) = R_s(\mathbf{x}_s, \mathbf{x}_s), \qquad \widetilde{\rho}_s(\mathbf{m}_s) = \widetilde{R}_s(\mathbf{m}_s, \mathbf{m}_s). \quad (2.9)$$

Seeing that $|\Psi|^2$ is assumed to be normalized to unity in the volume $V$, from (2.4) and (2.5) the normalization conditions follow

$$\int_V \rho_1(\mathbf{r}) d\mathbf{r} = N, \qquad \int_{(\infty)} \widetilde{\rho}_1(\mathbf{p}) d\mathbf{p} = N. \quad (2.10)$$

In a state of thermodynamic equilibrium, the reduced density matrices $R_s(\mathbf{x}_s, \mathbf{x}'_s)$ are given by Eq. (2.20) of [1] (see also Eq. (7.12) of [2]):

$$R_s(\mathbf{x}_s, \mathbf{x}'_s) = \frac{1}{2\pi i (2\pi\hbar)^{3s} s!} \int d\mathbf{m}_s \int_C dz\, n_s(z) v_s(\mathbf{x}_s, \mathbf{m}_s, z) \sum_P (\pm 1)^p \exp\left[\frac{i}{\hbar} \sum_{k=1}^{s} (\mathbf{p}_k \mathbf{r}_k - \mathbf{r}'_k \mathsf{P} \mathbf{p}_k)\right], \quad (2.11)$$

where the summation is over all permutations $P$ of the indices of the vector $\mathbf{p}_k$ in front of which the permutation operator $\mathsf{P}$ is placed, $p$ is the parity of the permutation, the contour of integration $C$ in the complex plane of $z$ is specified in [1]. Although we assume the particles to be spinless, for the purposes of illustration we accept the possibility of Bose statistics as well as Fermi. In (2.11) and in the following, the upper sign refers to bosons and the lower one to fermions. The function $v_s(\mathbf{x}_s,\mathbf{m}_s,z)$ is to be found from the equation

$$\frac{\hbar^2}{2m}\sum_{j=1}^{s}\nabla_j^2\, v_s(\mathbf{x}_s,\mathbf{m}_s,z) + \frac{i\hbar}{m}\sum_{j=1}^{s}\mathbf{p}_j\nabla_j\, v_s(\mathbf{x}_s,\mathbf{m}_s,z) + \left[z - \frac{1}{2m}\sum_{j=1}^{s}\mathbf{p}_j^2 - U_s(\mathbf{x}_s)\right]v_s(\mathbf{x}_s,\mathbf{m}_s,z) = 1. \quad (2.12)$$

The last equation of the hierarchy which links its $s$th and $(s+1)$th members is the one for the effective potentials $U_s(\mathbf{x}_s)$:

$$\rho_s(\mathbf{x}_s)\nabla_1 U_s(\mathbf{x}_s) = \rho_s(\mathbf{x}_s)\nabla_1\sum_{j=2}^{s}K(|\mathbf{r}_1-\mathbf{r}_j|) + \int\rho_{s+1}(\mathbf{x}_{s+1})\nabla_1 K(|\mathbf{r}_1-\mathbf{r}_{s+1}|)d\mathbf{r}_{s+1}. \quad (2.13)$$

The functions $n_s(z)$ in (2.11) are given by

$$n_s(z) = A_s e^{-z/\tau}, \qquad A_s = s!\,\rho^{s-1}\left(\frac{2\pi\hbar^2}{m\tau}\right)^{3(s-1)/2} A, \quad (2.14)$$

where the constant $A = A_1$ is fixed by the normalization of (2.10). As to the quantity $\tau$, it depends upon the temperature $\theta$ (here in units of energy) and the average particle density $\rho = N/V$; the interaction potential also plays a role. The function $\tau(\theta,\rho)$ is investigated in [1] (see also [2]).

If one puts (2.11) into (2.8), one will obtain equilibrium density matrices in momentum space. It is more instructive to consider the diagonal elements of the reduced density matrices of (2.9) that have a direct physical meaning. For example, the diagonal elements $\rho_s(\mathbf{x}_s)$ directly yield probabilities of specified configurations of the particles in space. If we put $\mathbf{x}_s = \mathbf{x}'_s$ in (2.11), the resulting expression can be cast in the form

$$\rho_s(\mathbf{x}_s) = \int W(\mathbf{x}_s,\mathbf{m}_s)d\mathbf{m}_s, \quad (2.15)$$

where

$$W(\mathbf{x}_s,\mathbf{m}_s) = \frac{1}{2\pi i(2\pi\hbar)^{3s}s!}\int_C dz\, n_s(z)v_s(\mathbf{x}_s,\mathbf{m}_s,z)\sum_P(\pm 1)^P \exp\left[\frac{i}{\hbar}\sum_{k=1}^{s}\mathbf{r}_k(\mathbf{p}_k-P\mathbf{p}_k)\right]. \quad (2.16)$$

To obtain $\tilde{\rho}_s(\mathbf{m}_s)$ we substitute (2.11) into (2.8) and set $\mathbf{m}_s = \mathbf{m}'_s$. In the integrand, one will have a $\delta$-function in view of (2.6), which enables one to carry out one of the integration. It is necessary, however, to permute the variables of integration in an appropriate way. In doing so account must be taken of the fact that the function $v_s(\mathbf{x}_s,\mathbf{m}_s,z)$ does not change if the coordinates and momenta are permuted simultaneously, which follows from (2.12) for the potentials $U_s(\mathbf{x}_s)$ are symmetric in the coordinates [1,2]. As a result we shall arrive at

$$\tilde{\rho}_s(\mathbf{m}_s) = \int W(\mathbf{x}_s,\mathbf{m}_s)d\mathbf{x}_s \quad (2.17)$$

with the same function $W(\mathbf{x}_s,\mathbf{m}_s)$ of (2.16). Therefore this last function plays the role of a *s*-particle Wigner function [5,6]. It gives the distribution over the positions when integrated over the momenta, and the distribution over the momenta when integrated over the coordinates.

Of special interest is the function $\widetilde{\rho}_1(\mathbf{p})$ that determines the overall distribution over the momenta in the system. From (2.17), (2.16) and (2.14) one has

$$\widetilde{\rho}_1(\mathbf{p}) = \frac{A}{(2\pi)^4 i\hbar^3} \int_V d\mathbf{r} \int_C dz\, e^{-z/\tau}\, v_1(\mathbf{r},\mathbf{p},z). \tag{2.18}$$

Concluding the section let us make a remark. The above consideration shows that the quantities $\mathbf{p}_k$ are momenta of the particles. In Ref. [2] it was stated erroneously that these quantities are not genuine momenta of the particles and become the genuine momenta solely in the classical limit. Only the use of the momentum representation allows one to establish correctly the physical meaning of the quantities $\mathbf{p}_k$ utilized in [1].

## 3. Distribution over momenta in a fluid

In the case of a uniform fluid, the function $v_1(\mathbf{r},\mathbf{p},z)$ does not depend upon $\mathbf{r}$ as well as the potential $U_1(\mathbf{r})$, and one can set $U_1 = 0$ [1]. If so, Eq. (2.12) yields at once

$$v_1(\mathbf{p},z) = \frac{1}{z - \dfrac{\mathbf{p}^2}{2m}}. \tag{3.1}$$

When (3.1) is inserted into (2.18), the integration over $z$ is readily carried out by the residue theorem with the result

$$\widetilde{\rho}_1(\mathbf{p}) = \frac{N}{(2\pi m\tau)^{3/2}} e^{-\mathbf{p}^2/(2m\tau)}, \tag{3.2}$$

into which the constant $A$ from Eq. (5.3) of [1] has been substituted (the value of $A$ can also be calculated from the normalization of (2.10)). This is a Maxwellian-type distribution (the distribution is not Maxwellian inasmuch as the parameter $\tau$ is not the temperature $\theta$). The result obtained is rather unexpected because it is valid for an ideal gas as well, while for the ideal gas one would expect a Bose or Fermi distribution.

This paradox can be resolved as follows. In Refs. [7, § 10.5] and [8, p. 209], the authors demonstrate that the ideal Bose gas in the ground state ($\theta = 0$) exhibits a pathology because its density is not constant in space and depends upon boundary conditions, which does not make physical sense and is even absurd. To remove the pathology it is necessary to take account of interparticle interactions [7,8]. A repulsive interaction, howsoever weak it may be, should smooth out any nonuniformity of the bulk density [7]. In quantum mechanics, however, coordinates and momenta cannot be independent of one another, which manifests itself in the uncertainty relations in particular. Once the interaction, be it extremely weak, drastically changes the distribution over the coordinates, it must drastically affect the distribution over the momenta as well. Although we have discussed the Bose gas at $\theta = 0$, the above reasoning should hold at $\theta \neq 0$ and for a Fermi gas, too. It is to be added also that there are no individual energy levels for the particles in interacting systems, whereas the Bose and Fermi distributions are formulated primarily as distributions in which just the individual energy levels figure.

As to the approach used in the present paper, its equations are valid in the thermodynamic limit ($V \to \infty$ and $N \to \infty$ with $N/V$ = constant) on the assumption that the interaction potential $K(|\mathbf{r}_j - \mathbf{r}_k|)$ is of arbitrary form [1,2]. To obtain an ideal gas it is sufficient to put $K(|\mathbf{r}_j - \mathbf{r}_k|) \equiv 0$. Owing to the thermodynamic limit the gas density will remain constant with the distribution over the momenta given by (3.2).

In this context another question arises. When considering the equation for the function $\tau(\theta,\rho)$ in [1] a condition on solutions of the equation was imposed according to which, in the limit of noninteracting particles, one should arrive at properties characteristic of ideal quantum gases. However, these same properties are obtained on a base of the Bose or Fermi distribution. Here one can proceed from the following. Even an extremely small variation of the interaction potential can substantially change the wavefunction $\Psi(\mathbf{x}_N,t)$ of the system while this should not tell upon macroscopic properties of the system (we do not even know the interaction potentials with absolute precision). Therefore, if we take the cases $K \equiv 0$ and $K \to 0$, the macroscopic (thermodynamic) properties should be identical even if the wavefunctions are different. It should be remarked that the system of noninteracting particles can never reach an equilibrium state. When one speaks of the thermodynamic quantities of such a system, one implies the most probable values which should remain the same if $K \equiv 0$ or $K \neq 0$ but $K \to 0$. It should be added also that, when treating the equation for $\tau(\theta,\rho)$ in [1], one does not use the Bose and Fermi distributions themselves, one resorts only to the expression for the internal energy $E$ of the ideal quantum gases (the expression used for the pressure $p$ follows from Eqs. (3.8) and (3.9) of [1] when $K \equiv 0$: $pV = 2E/3$).

We revert now to Eq. (3.2). Examples considered in [1] demonstrate that in the case of Bose systems the function $\tau(\theta,\rho)$ becomes negative at sufficiently low temperatures whereas it must be positive according to (3.2). It is clear that this inconsistency is due to a manifestation of Bose-Einstein condensation, and the approach used is to be modified to take this phenomenon into account, which is done in Ref. [9]. As to a Fermi system, the function $\tau(\theta,\rho)$ remains positive down to zero temperature. For an ideal gas of spinless fermions, $\tau = 2\varepsilon_0/5$ at $\theta = 0$, where $\varepsilon_0$ is the Fermi energy [1]. The same result remains valid for spin-half fermions although the Fermi energy is different [3].

We now turn to a superfluid. In this case, the density matrices break up into two parts, one of them describing the normal fraction while the other the condensate [9,10]. The part describing the normal fraction can be treated as above and leads to (3.2). The singlet density matrix describing the condensate is of the form (see Eq. (25) of [10])

$$R_1(\mathbf{r},\mathbf{r}') = \rho_c e^{i\frac{\mathbf{p}_0}{\hbar}(\mathbf{r}-\mathbf{r}')} u_1(\mathbf{r}) u_1^*(\mathbf{r}'). \tag{3.3}$$

For the superfluid, one is to take $u_1 = 1$. Substituting this into (2.8) and exploiting (2.6) (see also the remark that follows (2.6)) yields

$$\tilde{\rho}_1(\mathbf{p}) = N_c \delta(\mathbf{p} - \mathbf{p}_0), \tag{3.4}$$

where $N_c = \rho_c V$ is the number of particles in the condensate. Eq. (3.4) indicates that the particles of the condensate have a momentum equal to $\mathbf{p}_0$ although it is usually presumed that these particle have zero momentum.

A remark should be made at this point. When trying to determine the condensate density $\rho_c$ on a base of Eq. (3.4) with use made, for example, of neutron scattering [11], it should be kept in mind that, in a confined volume, the superflows must close upon themselves, so that they form a pattern with various directions of the vector $\mathbf{p}_0$ [9,12]. Special investigation is required to find out how the pattern manifests itself in the dynamic structure factor measured.

**4. Distribution over momenta in a crystal**

In the case of a crystal, the function $v_1(\mathbf{r},\mathbf{p},z)$ is of the form [13,14]

$$v_1(\mathbf{r},\mathbf{p},z) = \sum_{l,m,n=-\infty}^{\infty} b_{lmn}(\mathbf{p},z) e^{i\mathbf{A}\mathbf{r}}, \tag{4.1}$$

wherein $\mathbf{A} = l\mathbf{a}_1 + m\mathbf{a}_2 + n\mathbf{a}_3$ with the basic reciprocal-lattice vectors $\mathbf{a}_1$, $\mathbf{a}_2$ and $\mathbf{a}_3$. Upon substituting this into (2.18) and retaining only the term proportional to the volume $V$ (other terms oscillate when $V$ increases) one obtains

$$\tilde{\rho}_1(\mathbf{p}) = \frac{VA}{(2\pi)^4 i\hbar^3} \int_C dz\, e^{-z/\tau} b_0(\mathbf{p},z) \tag{4.2}$$

with $b_0(\mathbf{p},z) = b_{000}(\mathbf{p},z)$. The last function is given by Eq. (97) of [14], which enables one to integrate over $z$:

$$\tilde{\rho}_1(\mathbf{p}) = \frac{VA}{(2\pi\hbar)^3} v_0 \psi_0^2(\mathbf{p}) e^{-\frac{\varepsilon_0(\mathbf{p})}{\tau}}, \tag{4.3}$$

where $v_0$ is the volume of the elementary crystal cell, the functions $\psi_0(\mathbf{p})$ and $\varepsilon_0(\mathbf{p})$ being defined in Appendix A of [14] (cf. Eq. (98) of [14] with $B = A$). These last functions depend on the interaction potential and the crystalline structure. As distinct from Eq. (3.2) for a fluid, the distribution over momenta in the crystal is not universal.

As to a superfluid crystal, the function $u_1(\mathbf{r})$ that figures in (3.3) is now of the form [10,14]

$$u_1(\mathbf{r}) = \sum_{l,m,n=-\infty}^{\infty} c_{lmn}\, e^{i\mathbf{A}\mathbf{r}}. \tag{4.4}$$

We place this in (3.3) and analogously with (3.4) we obtain

$$\widetilde{\rho}_1(\mathbf{p}) = N_c \sum_{l,m,n=-\infty}^{\infty} |c_{lmn}|^2\, \delta(\mathbf{p} - \mathbf{p_0} - \hbar\mathbf{A}). \tag{4.5}$$

To verify this formula, one can calculate the momentum of the crystal $\mathbf{P}$ with use made of the formula in the following way:

$$\mathbf{P} = \int_{(\infty)} \mathbf{p}\, \widetilde{\rho}_1(\mathbf{p})\, d\mathbf{p} = N_c \left( \mathbf{p}_0 + \hbar \sum_{l,m,n=-\infty}^{\infty} \mathbf{A}\, |c_{lmn}|^2 \right), \tag{4.6}$$

where Eq. (35) of [10] has been utilized. The last expression fully coincides with Eq. (40) of [10] obtained in the position representation ($N = N_c$ in that equation).

From (4.5) we see that the momentum distribution in the crystal is rather curious. It consists of $\delta$-peaks that are arranged periodically in momentum space, the periodicity being characterized by the basic reciprocal-lattice vectors $\mathbf{a}_1$, $\mathbf{a}_2$ and $\mathbf{a}_3$ relevant to the crystal lattice, although the peaks are not identical because of the factor $|c_{lmn}|^2$. The peaks are displaced from the origin by the vector $\mathbf{p}_0$. Even if the condensate crystal is not superfluid ($\mathbf{p}_0 = 0$), the periodic structure in momentum space remains.

It should be observed that, when discussing the concept of the condensate in the concluding section of [14], the uncertainty principle was used incorrectly. It was presumed that each particle in the crystal could move only within its crystalline cell. In actual truth, the wavefunction does not vanish in the interstices, which indicates that the particles can quit their cells. This concerns especially the particles of the condensate that are in an exceptional state. Qualitative considerations cannot forecast their behaviour. Only strict mathematical calculations permit one to find out the momentum distribution for these particles, and the distribution turns out to be fairly unusual.

## 5. Conclusions

The present paper shows that the use of the momentum representation for the equilibrium reduced density matrices opens up new opportunities in studies of macroscopic quantum systems both nonsuperfluid and superfluid. We have found that the distribution over momenta in a quantum fluid is not a Bose or Fermi distribution even in the limit of practically noninteracting particles. The distribution looks like a Maxwellian one although, strictly speaking, it is not Maxwellian inasmuch as it depends not only on the temperature but also on the type of statistics, the density and the interaction potential. The momentum distribution in a quantum crystal is more complicated and can be found only upon solving equations describing the crystal. The momentum distribution in a superfluid contains a δ-function relevant to the condensate. Fairly curious is the momentum distribution for the condensate in a superfluid crystal. The distribution consists of δ-peaks that are arranged periodically in momentum space. The periodical structure remains if the condensate crystal is not superfluid.